\documentclass[aps,pra,amsmath,amssymb,twocolumn,showpacs]{revtex4}
\usepackage{amsmath,graphicx,float,epsfig,float,tabularx,multirow}
\usepackage{epstopdf,color,subfigure,pstricks}
\newcommand{\tr}[1]{\mathrm{Tr}\left[ {#1} \right]} 
 
\newcommand{\ket}[1]{\left\vert {#1} \right\rangle} 
\newcommand{\bra}[1]{\left\langle {#1} \right\vert} 
 
\def\a{\hat{a}_1}
\def\b{\hat{a}_2}
\def\ncd{{\rm T}} % symbol for the nonclassical depth
\def\Idop{\hat{\mathbb{I}}}
\newcommand{\STErev}[1]{{#1}}
\newtheorem{Def}{Definition}
\begin{document}
%%%%%%%%%%%%%%%%%%%%%%%%%%%%%%%%%
\title{Single- and two-mode quantumness at a beam splitter}
\author{Matteo Brunelli}\email{mbrunelli01@qub.ac.uk}
\affiliation{Centre for Theoretical Atomic, Molecular and Optical Physics, 
School of Mathematics and Physics, Queen's University Belfast, 
Belfast BT7\,1NN, United Kingdom }
\author{Claudia Benedetti}\email{claudia.benedetti@unimi.it}
\affiliation{Dipartimento di Fisica, Universit\`a degli Studi di 
Milano, I-20133 Milano, Italy}
\author{Stefano Olivares}\email{stefano.olivares@mi.infn.it}
\affiliation{Dipartimento di Fisica, Universit\`a degli Studi di 
Milano, I-20133 Milano, Italy }
\affiliation{CNISM UdR Milano Statale, I-20133 Milano, Italy}
\author{Alessandro Ferraro}\email{a.ferraro@qub.ac.uk}
\affiliation{Centre for Theoretical Atomic, Molecular and Optical Physics, 
School of Mathematics and Physics, Queen's University Belfast, 
Belfast BT7\,1NN, United Kingdom}
\author{Matteo G.~A.~Paris}\email{matteo.paris@fisica.unimi.it}
\affiliation{Dipartimento di Fisica, Universit\`a degli Studi di 
Milano, I-20133 Milano, Italy}
\affiliation{CNISM UdR Milano Statale, I-20133 Milano, Italy}
\date{\today}
%%%%%%%%%%%%%%%%%%%%%%%%%%%%%%%%%
\begin{abstract}
In the context of bipartite bosonic systems, two notions of classicality
of correlations can be defined: $P$-classicality, based on the
properties of the Glauber-Sudarshan $P$-function; and $C$-classicality,
based on the entropic quantum discord. It has been shown that these two
notions are maximally inequivalent in a static (metric) sense ---
as they coincide only on a set of states of zero measure. We extend and 
reinforce quantitatively this inequivalence by addressing the dynamical
relation between these types of non-classicality in a paradigmatic
quantum-optical setting: the linear mixing at a beam splitter of a
single-mode Gaussian state with a thermal reference state. Specifically,
we show that almost all $P$-classical input states generate outputs that
are not $C$-classical. Indeed, for the case of zero thermal reference
photons, the more $P$-classical resources at the input the less
$C$-classicality at the output. In addition, we show that the
$P$-classicality at the input --- as quantified by the non-classical
depth --- does instead determine quantitatively the potential of
generating output entanglement. This endows the non-classical depth with
a new operational interpretation: it gives the maximum number of thermal
reference photons that can be mixed at a beam splitter without
destroying the output entanglement.  
\end{abstract}
\maketitle
%%%%%%%%%%%%%%%%%%%%%%%%%%%%%%%%%
\section{Introduction}\label{s:intro}
Since the early days of quantum mechanics considerable efforts have 
been spent in establishing whether a given physical system possesses
genuinely quantum features.  
%devising systematical methods (techniques)
%As soon as quantum mechanics was developed, it became clear that some
%criterion was needed to establish when a given physical system possesses
%features without a classical analogue. 
As far as bosonic systems are
concerned, Wigner first attacked this problem introducing a quantum
analogue to the classical phase-space \cite{Wig}. Later on, a systematic
approach was finally developed in the framework of quantum optics, with
the introduction of various classes of quasi-probability distributions
defined over the quantum phase-space. Specifically, the analytical
features of such distributions unveil physical constraints: whenever the
normally-ordered distribution function --- called Glauber-Sudarshan
$P$-function \cite{gl1,sud1}--- behaves like a regular probability
distribution, the corresponding state can be described as a statistical
ensemble of classical fields and, in this sense, it cannot show any
non-classical feature \cite{mandel}. In the following, these states will
be referred to as $P$-classical states.  
\par
On the other hand, the more recent development of quantum information
theory promoted a reconsideration of the quantumness of physical systems
from an information-theoretical perspective. Since quantum systems can
be correlated in ways unaccessible to classical ones, the discrimination
between classical and nonclassical states of a given system is pursued
by studying the nature of the correlations among its subparts. In
particular, quantum entanglement accounts for quantum correlations that
may lead to the violation of local realism \cite{HHHH}. Moreover, even
separable (\textit{i.e.}, non-entangled) states have been recognized to
retain non-classical features, leading to the introduction of an
entropic measure, called quantum discord, to capture the quantum
features of correlations beyond entanglement \cite{Modi}. Following this
criterion, classical states can be defined as states with vanishing
discord, and we will refer to this notion as $C$-classicality.
\par
Although both acceptable and well-grounded, the two foregoing notions of
classicality have been shown to be radically different, indeed maximally
inequivalent, in the following sense: only a zero-measure set of states
is classical according to both criteria \cite{PvsI}. Besides embodying a
matter of fundamental interest, this conclusion is also relevant for
practical purposes, since it enlightens different resources in quantum
information processing \cite{C+:14}. However, such a characterization is based on
purely geometrical considerations and, as a consequence, it is
intrinsically static. In particular, the relation between $C$- and
$P$-classicality in common physical processes remains unclear. In
addition, a quantitative comparison between these two notions in terms
of their respective figures of merit still lacks.
\par
In this work, we address the above issues in the context of quantum
optics, whose description in term of the phase-space offers a natural
framework to develop a quantitative analysis \cite{cah69}. A
paradigmatic setting in quantum optics is constituted by Gaussian states
and operations, due to their relevance for quantum technologies and
their thorough theoretical characterization \cite{
oli:rev,GS3,FOP:05}. Specifically, we address
the dynamical relation of $P$- and $C$-nonclassical states arising from
the linear mixing of Gaussian states at a beam splitter. In this
setting, the availability of analytical expressions to quantify Gaussian
$P$-classicality --- in terms of the \textit{non-classical depth} ---
and Gaussian discord and entanglement, is crucial to work out a
quantitative comparison between the various notions of non-classicality.
\par
In particular, we consider the mixing of a generic Gaussian state with a
reference thermal state, and explore the relationships between the
classicality of the input state and the $P$- and $C$-classicality at the
output. Specifically, we show that almost all $P$-classical input states
generate output states that are not $C$-classical. Indeed, for the case
of zero thermal reference photons, the more $P$-classical resources at
the input the less $C$-classicality at the output. These findings
strengthen the inequivalence between $P$- and $C$-classicality by
quantitatively extending it to a process in which correlations are
dynamically generated, rather then statically analyzed as in
Ref.\cite{PvsI}. In addition, we show that the $P$-classicality at the
input does instead determine quantitatively the potential of generating
output entanglement. This endows the non-classical depth with a new
operational interpretation: it gives the maximum number of thermal
reference photons that can be mixed at a beam splitter without
destroying the output entanglement.  
\par
The paper is structured as follows. In Section \ref{s:gauss} we give a
brief account on Gaussian states and their phase-space representation,
focusing on their bilinear  interaction in linear optical devices. In
Section \ref{s:noncl} we review the two notions of nonclassicality and
establish the notation for Gaussian discord, non-classical depth and
entanglement used in the following. The reader familiar with the
foregoing topic can skip the respective sections. In  Section
\ref{ss:vac} we analyze in details the generation of $P$- and
$C$-nonclassicality by  mixing of a Gaussian state with the vacuum,
whereas in Section \ref{ss:the} we focus attention to the mixing with a
thermal state, also introducing the concept of effective
nonclassicality. Section \ref{s:out} closes the paper with some
concluding remarks.
%%%%%
\section{Linear mixing of Gaussian states}\label{s:gauss}
\STErev{The simplest bilinear interaction involving two bosonic field
modes described by the annihilation operators $\a$ and $\b$
(with $[\hat{a}_k, \hat{a}_k^{\dagger}]=\Idop$)
corresponds to the mode mixing and it is described by an effective
Hamiltonian of the form $H_I \propto (\a^{\dagger}\b+ \a \b^{\dagger}$).
This kind of interaction is very common in different quantum systems,
ranging from optical modes in linear optical devices \cite{WM}
to collective modes in ultracold atoms \cite{meystre}, opto-
and nano- mechanical oscillators \cite{woolley:08,xiang:10,PZ:12,AKM:14} and
superconducting resonators \cite{W+:04,chirolli:10}. For the sake of clarity,
in this paper we focus on the quantum optics realm and we address the
correlations properties of the two optical modes emerging from a beam splitter
(BS) when the input ones are excited in Gaussian states.}
\par
Gaussian states (GSs) are states with Gaussian Wigner functions
\cite{schum:86} and an exhaustive information about them is provided by
the knowledge of the first and second statistical moments of the
quadrature operators. Information about correlations is
contained in the second moment and from now on we set the first moments
to zero, without loss of generality. 
\STErev{Upon introducing the quadrature operators
$\hat{q}=(\hat{a}+\hat{a}^{\dagger})/\sqrt2$ and
$\hat{p}=(\hat{a}-\hat{a}^{\dagger})/(i\sqrt2)$,} the covariance
matrix (CM) of a single-mode GS of $\hat{a}$ is defined as $\left[
\boldsymbol{\sigma}\right]_{k l}=\frac12 \langle \{R_k,R_l\}
\rangle-\langle R_k \rangle\langle R_l \rangle$, with
$k,l=1,2$, being $\mathbf{R}^T=(R_1,R_2) \equiv (\hat{q},\hat{p})$ the
vector of the quadratures and $\{ \cdot,\cdot \} $ the anticommutator.
Now, the canonical commutation relations take the form $[R_k,R_l]=i\,
\omega_{kl}$, where $\omega_{kl} =(1- \delta_{kl})(-1)^{l}$ are the
entries of the is the $2 \times 2$ symplectic form
$\boldsymbol{\omega}$. 
The set of the eigenvalues $(q,p)\in\mathbb{R}^2$ of the position and
momentum-like operators, endowed with the symplectic form
$\boldsymbol{\omega}$, spans the real symplectic space
$\Gamma=(\mathbb{R}^2,\boldsymbol{\omega})$, 
which is referred to as the phase space of the mode $\hat{a}_1$. 
A single-mode GS may always be written as 
\begin{equation}
\label{gen1G}
\varrho_1(n_t,n_s)=S(r) \nu(n_t) S^\dagger(r)\,,
\end{equation}
where
$S(r)=\exp \left[\frac12 \left(ra^{\dagger 2} 
-r^*a^2\right)\right]$, $r \in \mathbb{C}$,
is the squeezing operator and
$$\nu(n_t)=(n_t + 1)^{-1}\left[ n_t/(n_t+1) \right]^{a^{\dagger} a}\,,$$ 
is a thermal state with $n_t$ average number of photons, 
the quantity $n_s=\sinh^2 |r|$ will be referred to as the
number of squeezed photons.
\par
\begin{figure}[h!] 
\centering
\includegraphics[width=0.7\columnwidth]{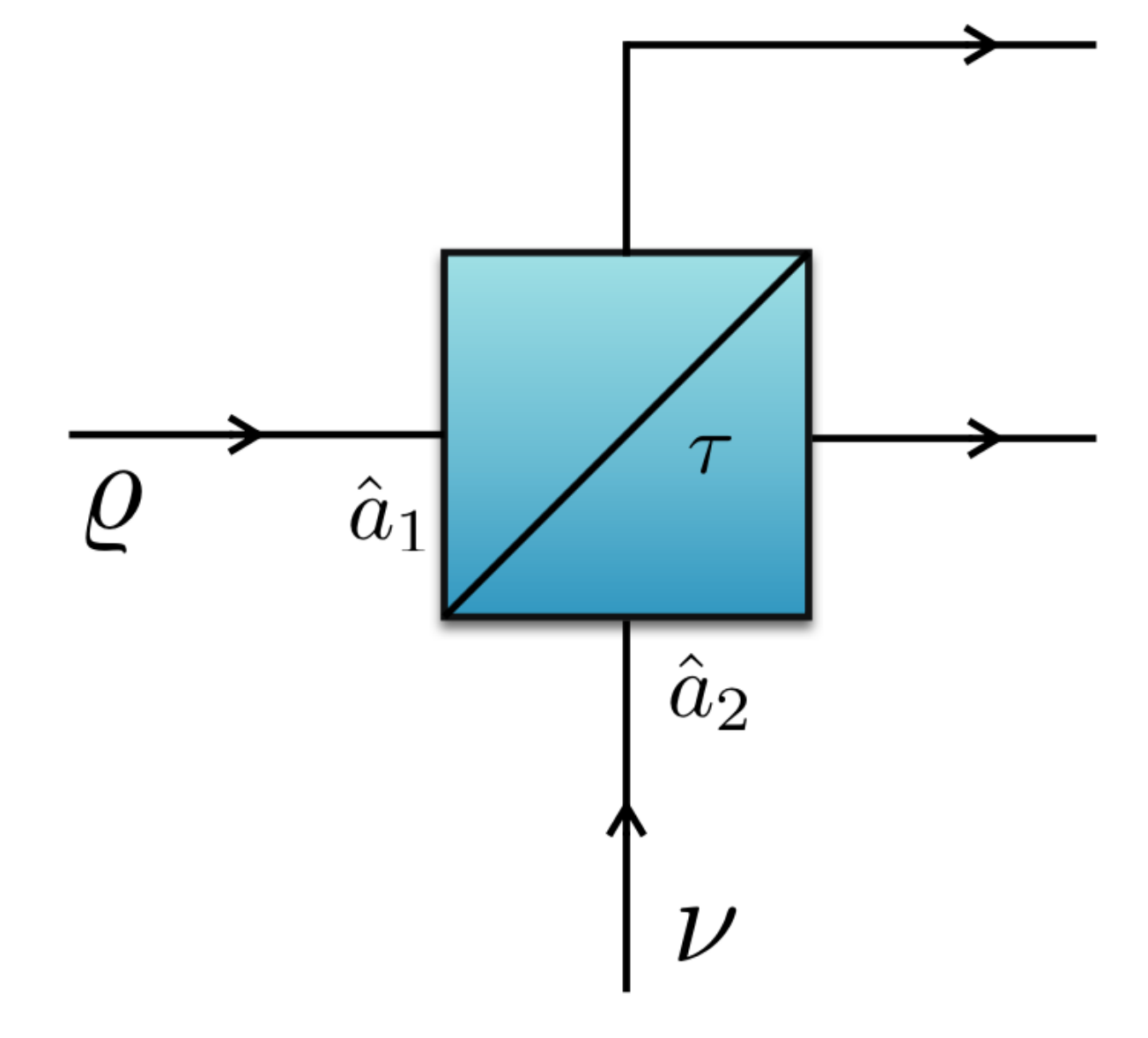} 
\caption{(Color online) 
Linear mixing of Gaussian states. The two input modes $\a$ and
$\b$, initially excited in the zero-mean Gaussian state
$\varrho=\varrho(n_s,n_t)$ and in the thermal state $\nu=\nu(n_2)$
respectively, enter a beam splitter of transmissivity $\tau$, after
which quantum correlations are eventually established.\label{f:BS}}
\end{figure}

\STErev{By choosing a suitable rotating frame, the lossless
BS exchange interaction is described by the unitary evolution
$U_\tau=\exp [ \theta (\hat{a}^{\dagger}_1\hat{a}_2 
- \hat{a}_1 \hat{a}^{\dagger}_2)]$,
with $\theta \in \mathbb{R}$ and where $\tau=\cos^2\theta$ denotes 
the transmissivity of the BS. If $\tau=1/2$, the BS is said to be balanced.
Being a bilinear interaction of modes,} this evolution preserves the
Gaussian character of the state, and in turn induces a symplectic
transformation $\mathsf{S}_{\tau}$ in the quantum phase space of the
composite system, namely
\begin{equation}\label{sympl}
\mathsf{S}_{\tau}=
\left(
\begin{array}{cc}
\sqrt{\tau } \, \mathbb{I} &  \sqrt{1-\tau } \, \mathbb{I} \\[1ex]
-\sqrt{1-\tau } \, \mathbb{I} &  \sqrt{\tau} \, \mathbb{I}
\end{array}
\right) \, ,
\end{equation} 
where $\mathbb{I}=\hbox{diag}(1,1)$.
Given two uncorrelated single-mode GSs, with CMs $\boldsymbol{\sigma}_1$ and 
$\boldsymbol{\sigma}_2$, respectively,
the symplectic transformation $\mathsf{S}_{\tau}$, acting by congruence
on the initial CM $\boldsymbol{\Sigma_0}= \boldsymbol{\sigma}_1 \oplus
\boldsymbol{\sigma}_2$, leads to the 
evolved CM $\boldsymbol{\Sigma}=\mathsf{S}_{\tau}\, \boldsymbol{\Sigma_0}\,
\mathsf{S}^T_{\tau}$.
As mentioned above and schematically depicted in Fig. \ref{f:BS}, in this work 
we will consider a bipartite quantum system of modes $\a$ and $\b$. The mode 
$\a$ is initially in  the zero-mean GS $\varrho=\varrho(n_s,n_t)$ while mode $\b$ is 
in a thermal state $\nu=\nu(n_2)$. Without loss of generality we will assume a real
squeezing parameter $r \in \mathbb{R}$. Using this parametrization, the average 
number of photons in the first mode $\langle\a^{\dagger}\a\rangle_{\varrho} \equiv
\hbox{Tr}[\a^{\dag}\a\,\varrho(n_s,n_t)]$ explicitly
reads:
\begin{equation}\label{toten}
\langle\a^{\dagger}\a\rangle_{\varrho} =
n_t + (1+ 2 n_t) n_s \equiv n_1.
\end{equation}
%%%
\par
%%%
In the phase space, the GSs $\varrho$ and $\nu$ are represented
by the $2\times 2$ CMs:
\begin{equation}\label{sigmarho}
\boldsymbol{\sigma}_{\varrho}=
\mathrm{diag}{\left(\frac12 + n_1 + \Delta, \frac12 + n_1 - \Delta
\right)} \, , \end{equation}
and
\begin{equation}\label{sigmanu}
\boldsymbol{\sigma}_{\nu}=
\left(\frac12 + n_2 \right) \mathbb{I} \, ,
\end{equation}
respectively, where $\Delta=(1+2 n_t) \sqrt{n_s (1+ n_s)}$.  Since the
initial state $R_0$ of the bipartite system is chosen to be factorized,
namely $R_0=\varrho\otimes\nu$, the total number of excitations is given
by $\langle\hat{a}^{\dagger}_1\hat{a}_1
+\hat{a}^{\dagger}_2\hat{a}_2\rangle_{R_0}=n_1+n_2\equiv
N \, .$ \STErev{Now we let the state $R_0$ evolve through a lossless BS
of transmissivity $\tau$ using the symplectic transformation
$\mathsf{S}_{\tau}$ given by Eq.~(\ref{sympl}).  The $4 \times 4$ CM of
the two-mode output state $R = U_\tau R_0 U_\tau^{\dag}$ is
$\boldsymbol{\Sigma}=\mathsf{S}_{\tau} (\boldsymbol{\sigma}_{\varrho}
\oplus 
\boldsymbol{\sigma}_{\nu}) \mathsf{S}^T_{\tau}$ and it reads:}
\begin{equation}\label{sigmamatrix}
\boldsymbol{\Sigma}=
\left(
\begin{array}{cccc}
 a_+  & 0 & c_+ & 0 \\
 0 & a_-  & 0 & c_- \\
 c_+ & 0 & b_+  & 0 \\
 0 & c_- & 0 & b_- 
\end{array}
\right) \, ,
\end{equation}
where:
\STErev{
\begin{subequations}\label{sigmaeq}
\begin{align}
a_{\pm} &=\left(\frac{1}{2}+n_2\right) 
(1-\tau )+\left(\frac{1}{2}+ n_1\pm \Delta \right)\tau \, ,
\\[1ex]
b_{\pm} &=\left(\frac{1}{2}+n_2\right) 
\tau+\left(\frac{1}{2}+n_1\pm \Delta\right) (1-\tau ) \, ,
\\[1ex]
c_{\pm} &=\left[\left(\frac{1}{2}+n_2\right)-
\left(\frac{1}{2}+n_1\pm \Delta \right)\right]
\sqrt{\tau(1-\tau)} \,.
\end{align}
\end{subequations}
}
It is useful to introduce the following local symplectic invariants,
which in this case are given by $I_1=a_+a_-, I_2=b_+b_-, I_3=c_+c_-$ and
$I_4=\det\boldsymbol{\Sigma}$.  Via symplectic
diagonalization $\boldsymbol{\sigma}_{AB}$ can be cast into the diagonal
form $\mathrm{diag}(\lambda_+,\lambda_+,\lambda_-,\lambda_-)$ where the
expression of the 
symplectic eigenvalues is given by \cite{ser:04}:
\begin{equation}
\lambda_{\pm}=\sqrt{\frac{I_1+I_2+2I_3\pm\sqrt{(I_1+I_2+2I_3)^2-4I_4}}{2}}\, .
\end{equation}
\STErev{Positivity of $\varrho_{AB}$ requires $\lambda_{-}\ge 1/2$.}
%%%%%
\section{Nonclassicality for bosonic systems}\label{s:noncl}
Let us now review in some details the concepts of non-classicality we
are going to consider, together with their respective figures of merit.
The reader familiar with those concepts can skip this Section.
\subsection{Nonclassicality in the phase space: P-classicality}
Any bipartite bosonic
state described by the density matrix $\varrho_{AB}$ can be always
expanded in terms of coherent states as follows:
\begin{equation}\label{pfunction}
\varrho_{AB}=
\int_{\mathbb{C}}d^2\alpha\int_{\mathbb{C}}
d^2\beta\,P(\alpha,\beta)\ket{\alpha}\bra{\alpha}
\otimes\ket{\beta}\bra{\beta} \, ,
\end{equation}
where $\ket{\alpha}$ and $\ket{\beta}$ are coherent states of the two
modes and $P(\alpha,\beta)$ is the Glauber-Sudarshan $P$-representation
of the state. $P(\alpha,\beta)$ provides a complete characterization of
the state.  Equation (\ref{pfunction}) suggests that the state of the
electromagnetic field can be regarded as a mixture of coherent states,
weighted by $P(\alpha,\beta)$. However, in general  the $P$-function
cannot be regarded as a probability density function. On the other hand,
when all the conditions for the $P$-function to be a probability density
are
satisfied, one can conclude that it describes a classical state of the
bosonic field, motivating the following definition:
\begin{Def}\label{P:def}
(\textit{$P$-classicality}) A state $\varrho_{AB}$ 
of a two-mode bosonic field is called $P$-classical if 
$P(\alpha,\beta)$ is a regular and normalized positive function.
\end{Def}
\STErev{In the case of a single-mode state $\varrho$, we can introduce a
generalized $s$-ordered Wigner function which encompasses all the 
quasi-probability distributions :}
\begin{equation}\label{sordered}
W_s(\alpha)=\int_{\mathbb{C}}{\frac{d^2
\lambda}{\pi^2}e^{\alpha \lambda^* - \alpha^* 
\lambda+(s/2)|\lambda|^2}\tr{D(\lambda)\varrho}},
\end{equation}
where $D(\alpha)\equiv \exp(\alpha \hat{a}^{\dagger}- \alpha^*\hat{a})$
is the displacement operator. In the case of $s=-1,0,1$ one recovers the
Husimi, Wigner, and Glauber-Sudarshan functions, respectively. The
latter, more than any other representation, can depart from being a
well-behaved probability density. In order to understand this fact, let
us observe that the $s$-ordered Wigner function of a state is related to
the $P$-function ($s=1$) of the same state through a Gaussian
convolution, namely: 
\begin{equation}\label{convP}
W_s(\alpha)=\frac{2}{\pi(1-s)}\int_{\mathbb{C}}d^2\beta
\, \exp\left\{ -\frac{2\vert\alpha-\beta\vert^2}{1-s}\right\}P(\beta) \, ,
\end{equation}
that can be seen as a smoothing operation. Given the $P$-function of the
state of interest, as the parameter $s$ moves towards $-1$, the
resulting distributions $W_s(\alpha)$ get smoother and smoother. Since
for $s=-1$ the Husimi $Q$-function is recovered, we obtain a continuous
interpolation between $P$- and $Q$-function, and we are  guaranteed that
this smoothing operation always succeeds in giving a true probability
distribution. 
\par 
Based on this, it is possible to define a quantitative measure of
$P$-nonclassicality, that is the non-classical depth of a quantum state
\cite{lee,lee2}. To this aim it is useful to introduce the parameter
$\ncd=(1-s)/2$. For $\ncd$ large enough, Eq.~(\ref{convP}) leads to a
$P$-classical state and the smoothing operation is referred to as
complete. If $\Omega$ denotes the set  of all $\ncd$ which give a
complete smoothing of the
initial the $P$-function, the non-classical depth is defined as
\begin{equation}\label{ncldepth}
\ncd_m=\inf_{\ncd\,\in\, \Omega}(\ncd)\,.
\end{equation}
The non-classical depth ranges from 0 for coherent states to 1 for 
Fock states \cite{tak02}, whereas for a 
single-mode GS $\varrho$ we have \cite{seraf:05}:
\begin{equation}
\ncd_m= \max\left[\frac{1}{2}\left(1-\frac{e^{-2r}}{\mu} \right),0\right] ,
\end{equation}
where $\mu=\hbox{Tr}[\varrho^2]$ is the purity of the state and $r$ is
the squeezing parameter introduced in Eq. (\ref{gen1G}).
%%%
\subsection{Nonclassicality of correlations: C-classicality}
The
total amount of correlations between two classical systems $A$ and $B$ is
quantified by the mutual information $I(A:B)=H(A)+H(B)-H(A,B)$ where
$H(X)=-\sum_ x p_X(x)\log p_X(x)$ is the Shannon entropy of the random
variable $X$. By exploiting the relation $p_{AB}(a,b)=p_A(a\vert b)p_B(b)$ 
one also gets the equivalent expression of the mutual information $I(A:B)=
H(A)-H(A\vert B)$ in terms of the conditional entropy $H(A\vert B)=-\sum_{a,b}
p_B(b)\, p_A(a\vert b)\log p_A(a\vert b)$.   
The first expression of the mutual information
has an immediate extension to quantum systems, simply by replacing the
Shannon entropy with the Von Neumann entropy $S[\varrho]=-\tr{\varrho
\log \varrho}$, namely
$I_M(\varrho_{AB})=S(\varrho_A)+S(\varrho_B)-S(\varrho_{AB})$; if we
address GSs, it can be expressed as \cite{ser:04}
\begin{equation}
I_M(\varrho_{AB})=f\left(\sqrt{I_1}\right)+f\left(\sqrt{I_2}\right)
-f\left(\lambda_+\right)-f\left(\lambda_-\right),
\end{equation}
where $f(x)=(x+\frac12)\log(x+\frac12)-(x-\frac12)\log(x-\frac12)$. On
the other hand, the extension of the second expression to the quantum
realm involves a measurement on one of the two parties, say $B$,
described by the positive operation-valued measure (POVM)
$\{\Pi_k\}, \; \Pi_k\ge0,\; \sum_k \Pi_k = \Idop$. The probability to
obtain the outcome $k$ is given, according to the Born rule, by
$p_k=\hbox{Tr}[\varrho_{AB}\, \Idop\otimes\Pi_k]$ and the conditional
state of $A$ with respect to the $k$ outcome
$\varrho^{\Pi_k}_{A|B}=(p_k)^{-1} \hbox{Tr}_{B}[\varrho_{AB}\,
\Idop\otimes\Pi_k]$. The maximum amount of information we can gain on
the part $A$ by locally measuring the other part thus has the
non-trivial expression
\begin{equation}
\mathcal{C}_{A|B}(\varrho_{AB})=
\max_{\{\Pi_k\}}\left\{S\left(\varrho_A\right)
-\sum_k p_k S\left(\varrho^{\Pi_k}_{A|B}\right) \right\} \,,
\label{maxx}
\end{equation} 
and invokes  an optimization procedure over the set of all measurements.
The quantum discord is properly defined as the difference between these
two quantities \cite{zurek} 
\begin{equation}
\mathcal{D}_{A|B}(\varrho_{AB})=I_M(\varrho_{AB})-\mathcal{C}_{A|B}(\varrho_{AB})\,.
\end{equation}
We can thus conclude that a system shows some quantumness as soon as the
discord is different from zero, providing us with the following
criterion:
\begin{Def}\label{C:def}
\textit{($C$-classicality)} A state 
$\varrho_{AB}$ of a two-mode bosonic field is called $C$-classical if  
$\mathcal{D}_{A|B}(\varrho_{AB})=\mathcal{D}_{B|A}(\varrho_{AB})\equiv 0$.
\end{Def}
%%%%
If we restrict to the subclass of GSs and Gaussian
measurements, an analytical expression of the quantum discord can be
derived, which is  called Gaussian discord, and is given by
\begin{equation}\label{discordgauss}
\mathcal{D}_{A|B}(\varrho_{AB})=
f\left(\sqrt{E^{\mathrm{min}}_{A|B}}\right)
+f\left(\sqrt{I_2}\right)-f\left(\lambda_+\right)-
f\left(\lambda_-\right) \, ,
\end{equation}
where $E^{\mathrm{min}}_{A|B}$ has an analytical expression as a
function of the local symplectic invariants $I_k$
\cite{gio10,ade10,gu12,mas12,bla12}. It is worth noting that
for a large class of Gaussian states \cite{gea14} the Gaussian
discord of Eq. (\ref{discordgauss}) coincides with the quantum discord,
i.e. the maximum in Eq. (\ref{maxx}) is achieved by a Gaussian
measurement.
\par
A stronger form of quantum correlations with respect to discord is given
by quantum entanglement. In the case of two-mode GSs
a necessary and sufficient condition can be derived to assess the
presence of entanglement \cite{sim:00}. It is essentially based on the
positivity of $\varrho_{AB}$  under partial transposition (PPT), that is
the positivity of the density matrix obtained by the transposition
applied only to one part of a system \cite{per:96}. One can show in fact
that the symplectic eigenvalues of the partially transposed states are
given by:
\begin{equation}
\tilde{\lambda}_{\pm}=\sqrt{\frac{I_1+I_2-2I_3\pm
\sqrt{(I_1+I_2-2I_3)^2-4I_4}}{2}}\,,
\end{equation}
and a Gaussian state $\varrho_{AB}$ is entangled if and only if  $\tilde{\lambda}_- < 1/2$. In fact, 
a measure of entanglement is given by the logarithmic negativity \cite{vid:02}, that is
$E(\boldsymbol{\sigma})=\mathrm{max}\left[-\log(2\tilde{\lambda}_-),0\right]$.
%%%
\section{Non-classicality arising from mixing a Gaussian state 
with the vacuum}
\label{ss:vac}
We start considering the case in which the reference input state
of mode $\b$ is the vacuum, i.e. $n_2=0$ and, in particular
$\boldsymbol{\sigma}_{\nu}\rightarrow \boldsymbol{\sigma}_0\equiv
\frac12 \mathbb{I}$ and $n_1 = N$. 
%%%
\subsection{P-classicality}
The initial state of the system $R_0$ is clearly $C$-classical. On the
contrary, $P$-nonclassicality has to be addressed both in the input and
output channels, in order to see wether differences arise. The
Glauber-Sudarshan $P$-representation of $R_0$ is given by 
\begin{equation}\label{R0}
R_0=\int_{\mathbb{C}}d^2\alpha\int_{\mathbb{C}}d^2\beta\,
P_{\varrho}(\alpha)\,P_{0}(\beta)
\ket{\alpha}\bra{\alpha}\otimes\ket{\beta}\bra{\beta} \, ,
\end{equation}
where the $P$-function $P_{R_0}$ is factorized in the product of
$P_{\varrho}$ and $P_{0}$, being the latter the Glauber-Sudarshan
P-function of the vacuum.  Moreover, since $P_{0}$ is a well-behaved
probability density, any possible $P$-nonclassical feature of the input
state is due to the pathological behavior of $P_{\varrho}$ alone, and is
quantified by the nonclassical depth of Eq. (\ref{ncldepth}).  Since the
action of the BS evolution on the two-mode displacement operator is
$U_\tau\,D_a(\alpha)\otimes D_b(\beta)\,
U_\tau^{\dagger}=D_a(\sqrt{\tau}\alpha+\sqrt{1-\tau}\beta) 
D_b(\sqrt{\tau}\beta-\sqrt{1-\tau}\alpha)$, and it amounts to a rotation 
of the arguments, one obtains the following $P$-representation of $R$:
\begin{align}\label{Rrotate}
P_{R} (\alpha,\beta)=
P_{\varrho}(\sqrt{\tau}\alpha-\sqrt{1-\tau}\beta)\,
P_{0}(\sqrt{1-\tau}\alpha+\sqrt{\tau}\beta)\, .
\end{align}
It is apparent that the effect of the evolution only amounts to a
re-parametrization of the argument, that does not affect the functional
form of both $P_{\varrho}$ and $P_{0}$.  Thus we can conclude that
output state $R$ is $P$-nonclassical if and only if $\varrho$ is
$P$-nonclassical, that is, for our configuration \textit{the two-mode
$P$-nonclassicality of the output equals single-mode $P-$nonclassicality
of the input}. Let us remark, for the sake of
Sec.\ref{ss:the}, that the foregoing argument applies as well for the
case of mixing with a reference thermal state with positive temperature,
given that the P-function of a general thermal state is a well-behaved
probability density.
%%%
The non-classical depth in Eq.~(\ref{ncldepth})
relative to the mode $\a$, can be expressed as:
\begin{equation}\label{NCdepthmax}
\ncd_m=\mathrm{max}\left[\frac{1-2u}{2},0\right] \, ,
\end{equation}
where $u=\frac{1}{2}+n_1-\Delta$ is the minimum eigenvalue of the CM
$\boldsymbol{\sigma}_{\varrho}$, as it is apparent from
Eq.~(\ref{sigmarho}). The condition $\ncd_m=0$ singles out a
$P$-classicality threshold, which can be made explicit either as a
function of the thermal component $n_t$, hence having
\begin{equation}\label{NsNCl}
n_s^{\mathsf{P}}=\frac{n_t^2}{1+2 n_t} \, ,
\end{equation}
or of the squeezed ones
\begin{equation}\label{NthNCl}
n_t^{\mathsf{P}}=n_s+\sqrt{n_s(1+ n_s)} \, .
\end{equation}
Whenever the average number of squeezed photons exceeds
$n_s^{\mathsf{P}}$, or the thermal component falls below
$n_t^{\mathsf{P}}$ the state $\varrho$ turns out to be $P$-nonclassical.
From now on, otherwise differently stated, $n_s^{\mathsf{P}}$ shall be
employed as the \textit{$P$-classicality threshold}.
%%%%
\begin{figure}[h!] 
\centering
\includegraphics[width=0.9\columnwidth]{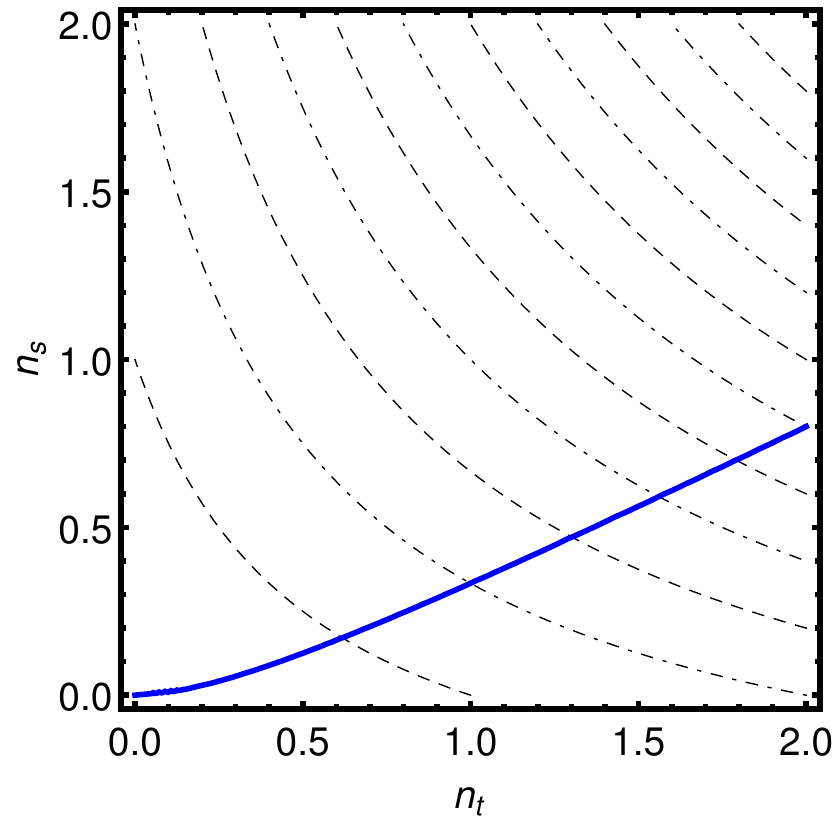} 
\caption{(Color online) 
Nonclassicality by mixing with the vacuum:
Plot of the implicit function T$_m=0$, which defines the
$P$-classicality threshold. It takes the explicit expression of Eq.
(\ref{NsNCl}) as a function of $n_t$, while as a function of $n_s$ it is
given by Eq. (\ref{NthNCl}). In this case the non-classicality threshold
$n_s^{\mathsf{P}}$ and the separability threshold $n_s^{\mathsf{sep}}$
coincide. The black lines are curves of fixed total energy $N=n_s
+n_t+2n_sn_t$ : dashed for odd values $N=2k+1\, ,\, k=0,1,\ldots$ and
dot-dashed for even values $N=2k ,\, k=1,2,\ldots$ of the total
energy.\label{f:NCthr}}
\end{figure}
%%%%
\subsection{C-classicality and generation of Gaussian discord}
We now address general quantum correlations, and investigate the
generation of Gaussian discord. Being the parties involved the output
modes $\hat{b}_1=U^{\dagger}\,\hat{a}_1\,U\,$ and
$\hat{b}_2=U^{\dagger}\,\hat{a}_2\,U\,$, we will refer to
$\mathcal{D}_{1|2}$ as the $b_1$-discord and $\mathcal{D}_{2|1}$ to the
$b_2$-discord; when they both coincide the symbol $\mathcal{D}$ will be
employed employed. In fact, this is the case of a balanced BS, namely
$\mathcal{D}_{1|2}(n_s,n_t,1/2)=\mathcal{D}_{2|1}(n_s,n_t,1/2) \quad
\forall \; n_s, \, n_t$.
\par
\begin{figure}[h!] 
\includegraphics[width=0.9\columnwidth]{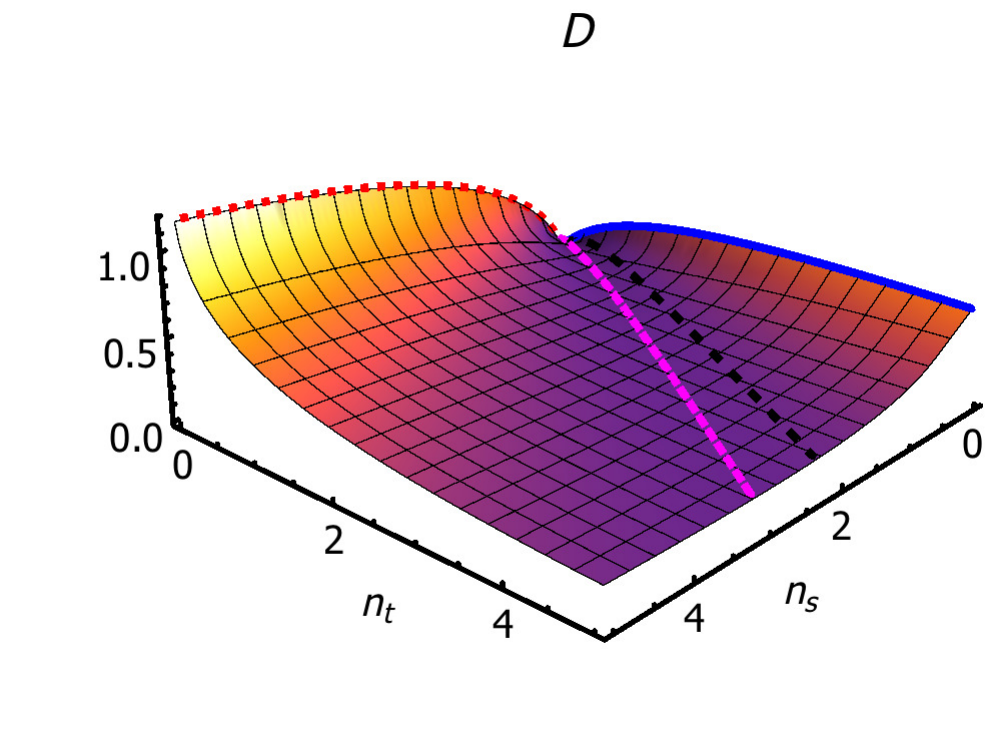} 
\includegraphics[width=0.95\columnwidth]{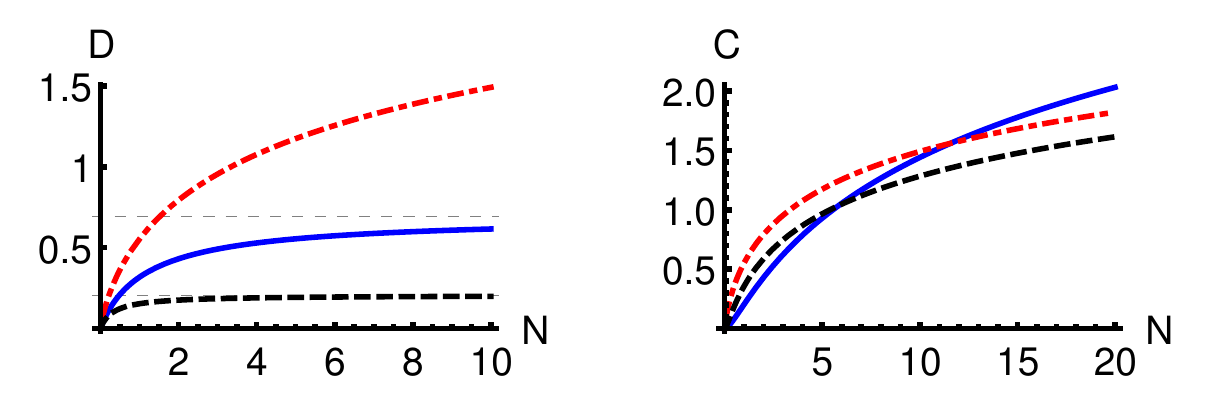} 
\caption{(Color online) 
Nonclassicality by mixing with the vacuum: In the upper panel 
we show the discord $\mathcal{D}$ as a function of the number of
squeezed and thermal photons $n_s$, $n_t$  for a balanced BS 
$\tau=1/2$. The dotted red line corresponds to the case of 
a squeezed vacuum state entering the beam splitter. The solid 
blue line corresponds to a thermal input
state, while the dashed black curve is the discord at the
$P$-classicality threshold. Finally the dashed magenta curve points out
the minimum value of the discord, obtained via numerical maximization.
The lower left panel shows the discord  as a
function of the total energy $N$; solid blue for thermal state, dashed
red for squeezed vacuum, black dot-dashed for $P$-classicality
threshold. The right panel shows the classical correlations for
the same input states and with the same color codes.
\label{f:Discord3D}} 
\end{figure}

In Fig. \ref{f:Discord3D} we show a plot of the Gaussian discord as a
function of the squeezed and thermal component of the input state
$\varrho$, for the balanced case $\tau=1/2$.  Except for the trivial
case of a vacuum input state $\varrho=\ket{0}\bra{0}$, it is apparent
that the discord is always positive and therefore, \textit{contrary to
the case of $P$-classicality, there is no $C$-classicality threshold}:
whatever the input state, a balanced BS is capable of generating quantum
correlations. This is also in agreement with the fact that typically
almost all states possess positive discord \cite{almostall}.
\par
From Fig.~\ref{f:Discord3D} a quantitatively relevant feature emerges.
Considering input states below the $P$-classicality threshold (denoted
by the dashed black curve), one can observe that the output discord and
hence the \textit{$C$-nonclassicality increases as the input
non-classical resources decreases}. This is true regardless the
constraints that one considers: either moving along the curves at
constant $n_s$, $n_t$, or total energy $N$, the discord increases as
$n_s$ decreases or $n_t$ increases. This is a quantitative feature that
strikingly confirms --- together with the absence of a $C$-classicality
threshold --- the inequivalence between the two notions of classicality
considered here.
\par
In Fig. \ref{f:Discord3D} the Gaussian discord corresponding to three
families of input states has been highlighted: the discord generated by
a thermal input state $\mathcal{D}^{\,\mathrm{th}}=\mathcal{D}(0,n_t)$
corresponds to the solid blue curve; the dotted red line is obtained
when the input state is the squeezed vacuum state, i.e.
$\mathcal{D}^{\,\mathrm{sq}} =\mathcal{D}(n_s,0)$; and the black dashed
line represents the value of the discord at the $P$-classicality
threshold, i.e.  $\mathcal{D}^{\,\mathsf{P}}=
\mathcal{D}(n_s^{\mathsf{P}},n_t)$.  For
these limiting cases, analytical expressions of the discord in terms of
the total energy $N$ of are available, even if quite cumbersome, and
hence not reported. Being functions of  a single quantity, they are
suitable for comparison and have been plotted in the lower panel of
Fig.~\ref{f:Discord3D}, together with the relative values of the
classical correlations  $C=I_M-\mathcal{D}$. Moreover, in Fig.
\ref{f:Discord3D} we also show, by a dashed magenta line, the curve
corresponding to the minimum value attained by the discord (for fixed $n_t$) 
obtained via numerical minimization.
\par
From the left bottom panel in Fig.~\ref{f:Discord3D}, we can see that
the discord is a monotonically increasing function of the total energy
$N$.  The discord saturates to a finite value both for a thermal
input state, for which we find
$$\lim_{N\rightarrow\infty}\mathcal{D}^{\,\mathrm{th}}=\log 2\,,$$ 
and at the $P$-classicality threshold \cite{caz13}, where 
\begin{align}
\lim_{N\rightarrow\infty}\mathcal{D}^{\,\mathsf{P}}= \frac12 \log\left(3+2
\sqrt{2}\right)-\frac32 \log\sqrt{2}\approx 0.2067\,.\label{dinf}\end{align}
Again we see that, for a fixed value
of the total energy, a thermal input state results in more quantum
correlations than a state lying on the non-separability boundary,
although in the latter case squeezing is involved. Actually, as we can
see from Fig. \ref{f:Discord3D}, the states corresponding to the
$P$-classicality threshold do not correspond to the states with minimum
output discord (dashed magenta curve), confirming again the
inequivalence between $P$- and $C$-classicality. The minimum output
discord curve has been obtained  numerically and we have not found
any clear physical picture of the class of states for which this minimum
is achieved.  On the other hand, for a squeezed vacuum state the discord
grows logarithmically for large $N$ values.  As a final remark, from the
right plot in the lower panel of Fig.~\ref{f:Discord3D}, it is apparent
that as the energy increases the classical correlations always increase
indefinitely, whereas as said, below the $P$-classicality threshold
discord is bounded. This behavior will become clear in the next
section, where entanglement will be considered.
\par
\begin{figure}[h!] 
\centering 
\includegraphics[width=0.85\columnwidth]{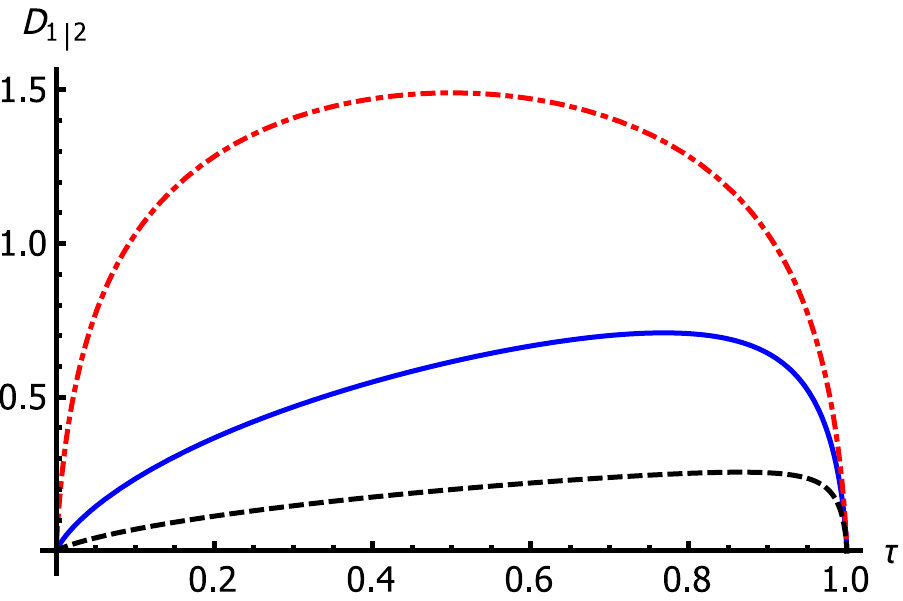}
\caption{(Color online) 
Nonclassicality by mixing with the vacuum: plot of the 
discord $\mathcal{D}_{1|2}(R)$ as a function of
the transmissivity $\tau$ for a given total energy $N=10$. Colors as in
Fig. \ref{f:Discord3D} . \label{f:Disctau}}
\end{figure}

If we now release the restriction of a balanced BS and inquire the
behavior of the discord with respect to $\tau$, we see that
$\mathcal{D}_{1|2}(R)$ and $\mathcal{D}_{2|1}(R)$  differ from each
others; for a generic value $\tau$ of the transmissivity, $b_1$-discord
and $b_2$-discord are simply related by an exchange of  the symplectic
invariants of $I_1$ and $I_2$, which amounts to a swap of the BS
transmissivity from $\tau$ to $1-\tau$.  In Fig. \ref{f:Disctau}, the
$b_1$-discord has been plotted as a function of $\tau$, for the relevant
cases already mentioned. Being obtained by the exchange of
transmissivity and reflectivity, the $b_2$-discord is simply given by a
reflection about the axis $\tau=\frac12$. Of course, in the limiting
cases of transmissivity 0 and 1, the discord falls to zero. Furthermore,
apart for a squeezed vacuum input state, the behavior of
$\mathcal{D}_{1|2}(R)$ is not symmetric with respect to $\tau=\frac12$.
By increasing the incoming energy, the maximum of
$\mathcal{D}_{1|2}(R)\; (\;\mathcal{D}_{2|1}(R)\;)$, and hence the
optima transmissivity, shifts towards $\tau=1\, (\,0\,)$. 
%%%
\par
%%%%
\subsection{Generation of Gaussian entanglement}
Let us now focus on the generation of Gaussian entanglement. 
The explicit expressions of the symplectic 
invariants are given by
\STErev{
\begin{subequations}\label{symplinv}
\begin{align}
I_1&= \frac{1}{4}+ n_t(1+n_t)\tau + N (1-\tau) \tau \,,\\[1ex]
I_2&= \frac{1}{4}+ n_t(1+n_t) (1-\tau) + N  (1-\tau)\tau \,,\\[1ex]
I_3&=-\left[(1-n_t)n_t + N \right](1-\tau)\tau\,,\\[1ex]
I_4&=\frac{1}{16}\left(1+2 n_t \right)^2 \, .
\end{align}
\end{subequations} 
By solving the equation $\tilde{\lambda}_-(n_s,n_t,\tau)=\frac12$ with
respect to $n_s$, one finds an analytic expression for the the number of
squeezing photons at the separability threshold:}
\begin{equation}\label{NsSep}
n_s^{\mathsf{sep}}=\frac{n_t^2}{1+2 n_t} \, ,
\end{equation}
which does not depend on $\tau$ and, most important, it equals the
$P$-classicality threshold Eq.~(\ref{NsNCl}). This is in agreement with
the general fact that $P$-nonclassicality is necessary and sufficient
for the generation of entanglement at a BS, regardless the
Gaussian nature of the input state \cite{msk02,xia02,WEP:03,asb05,oli09,
oli11,JLC:13,VS:14}. 
\par 
We are now going to analyze more in details the relationship between the
generation of Gaussian discord and entanglement. Since, although
analytical, the expression of the Gaussian discord is far too involved,
being in particular non invertible, we proceed in our analysis by
randomly sampling a large number of input Gaussian states and making
them evolve trough a BS of random transmissivity $\tau\in [0,1]$; for
each of them, minimum symplectic eigenvalue
of the partially transposed CM and Gaussian discord are then computed.
\begin{figure}[h!] 
\centering 
\includegraphics[width=.9\columnwidth]{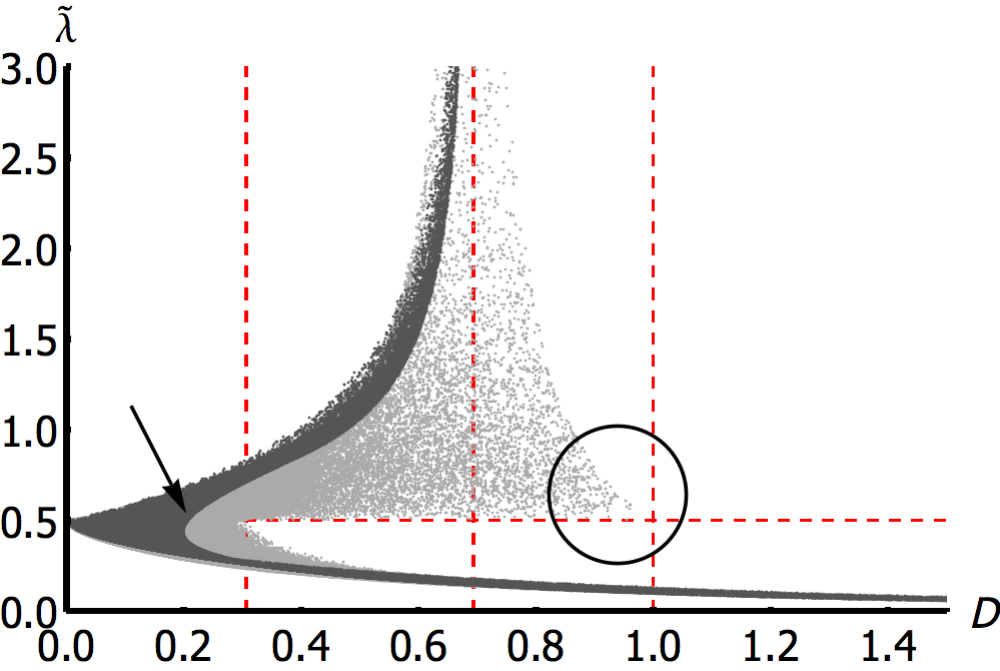}
\caption{(Color online) 
Nonclassicality by mixing with the vacuum: 
plot of the minimum symplectic eigenvalue of the partial
transpose CM $\tilde{\lambda}_-$ versus the discord
$\mathcal{D}_{1|2}(R)$ for randomly generated input states
$\varrho(n_s,n_t)$ evolving through a balanced BS (dark gray points),
and for random values of the transmissivity $\tau$ (light gray points).
The vertical dashed lines correspond to $1-\log2\approx 0.3069$,
$\log2\approx0.6931$, which is an asymptotic value for thermal states
entering a balance BS, and $1$, beyond which only entangled states
($\tilde{\lambda}_-<$  1/2) can be found. The black circle stresses the
portion of the plane occupied by states attaining the maximum value of
the discord, still being separable; the black arrow points the maximum
value of the discord achieved at the separability threshold
$\tilde{\lambda}_-$ = 1/2 in the case of a balanced BS.
\label{f:RndDiscVac}}
\end{figure}
\par
The results are shown in Fig.~\ref{f:RndDiscVac} (light gray points),
together with the plot obtained for evolutions trough a balanced BS
(dark gray points). Inspecting the latter distribution it is easy to
recover all the features already addressed in Fig. \ref{f:Discord3D}. In
particular, since for input thermal states of large energy the discord
has been found to reach the limiting value $\log2\approx0.6931$, the
distribution displays an asymptote, so that we can conclude that the
region of high $\tilde{\lambda}$ corresponds to highly excited input
thermal states. Moreover, as it can be seen following the dashed black 
line in Fig. \ref{f:Discord3D}, by moving on the separability threshold, 
i.e. considering the points laying on the line $\tilde{\lambda}_- = 1/2$, we
move from zero discord to the asymptotic value of Eq. (\ref{dinf}), 
which is obtained for infinite input energy and pointed out by 
the black arrow.
It is also possible to note that the minimum value of the discord is
attained slightly below $\tilde{\lambda}_- = 1/2$, as shown by the
dashed magenta line of Fig. \ref{f:Discord3D}.  
\par
More in general, considering arbitrary transmissivity (light gray points
in Fig.~\ref{f:RndDiscVac}) if the evolved state has a discord
$\mathcal{D}_{1|2}(R)>1$, it will be necessarily entangled
\cite{gio10,ade10}: the avoided region of the plane
$\{(\mathcal{D},\tilde{\lambda}_-)\; \vert\; \mathcal{D}>1\;,
\tilde{\lambda}_- > 1/2 \}$ shows that the discord for separable states
is always smaller than one.  It means that for separable---and hence
$P$-classical---input states, whatever the transmissivity, the discord
between the output modes cannot grow indefinitely~\cite{ade10}, by simply pumping
more energy. On the other hand, in the region
$0\le\mathcal{D}_{1|2}(R)\le1$, both entangled and separable states are
present. Another region of interest is the entangled region, namely
$\{(\mathcal{D},\tilde{\lambda}_-)\; \vert\;  \tilde{\lambda}_- \le
1/2 \}$, where the random generated points get ``horn-shaped". One
remarkable feature is that, when $\tilde{\lambda}_-$  approaches zero,
i.e. entanglement is high, the discord becomes, loosely speaking, nearly
a function of $\tilde{\lambda}_-$, and hence of entanglement itself.
\STErev{ In this case, we note that for $\mathcal{D}\gtrsim1$ the extent
of the region in Fig.~\ref{f:RndDiscVac} is bounded by two convergent
quantities. To better clarify this point we set $\tau=1/2$ and consider
an input squeezed vacuum state. For large $N\gg 1$ the analytic
expressions of the discord  $\mathcal{D}^{\,\mathrm{sq}}$ reads (at the leading
order):
\begin{equation}
\mathcal{D}^{\,\mathrm{sq}} \approx \log\left( \frac{\sqrt{N}}{2}\right)+1  \, ,
\end{equation}
while the minimum symplectic  eigenvalue of the partial transpose
$\tilde{\lambda}_-^{\,\mathrm{sq}}$ is 
\begin{equation}
\tilde{\lambda}_-^{\,\mathrm{sq}} \approx \frac{1}{4\sqrt{N}} \, 
\end{equation}
respectively. Form the previous equations it follows that in this limit
$\tilde{\lambda}_-^{\,\mathrm{sq}} \approx e^{1-\mathcal{D}^{\,\mathrm{sq}}}/8$, valid
for high discord value.} It turns out to be an upper bound for the
random distribution of symplectic eigenvalues in the entangled region, and
hence will be denoted as $\tilde{\lambda}_-^M$. Upon omitting the superscript
$\mathrm{sq}$, we may write 
\begin{equation}\label{lambdaM}
\tilde{\lambda}_-\,\le\,\tilde{\lambda}_-^{M}\,
\approx\, \frac{e^{1-\mathcal{D}}}{8} \qquad
\mathrm{for}\quad \mathcal{D}\gtrsim1
\end{equation}
\STErev{Analogously, our numerical analysis in the same region shows
that} the random generated points are always bounded from below
by $\tilde{\lambda}_-^m$, whose expression is
\begin{equation}\label{lambdam}
\tilde{\lambda}_-\,\ge\,\tilde{\lambda}_-^m\,
\approx \,\frac{e^{-\mathcal{D}}}{4} \qquad\mathrm{for}
\quad \mathcal{D}\gtrsim1\, .
\end{equation}
Putting together Eqs.~(\ref{lambdaM}) and (\ref{lambdam}) we conclude
that, for a fixed value of the discord $\mathcal{D}\gtrsim1$, the
distribution of minimum symplectic eigenvalues of the partially
transpose CM, is constrained in the range
\begin{equation}
\tilde{\lambda}_-^m \le \tilde{\lambda}_- \le \tilde{\lambda}_-^M \, .
\end{equation}
Therefore, for $\mathcal{D}\gg1$, $\,\tilde{\lambda}_-$ is an
exponentially decreasing function of the Gaussian discord.  
\par 
Particularly interesting is finally the region corresponding to highly
discordant---yet separable---states, stressed by a circle in
Fig.~\ref{f:RndDiscVac}. These are states sharing the maximum amount of
quantum correlations without invoking entanglement. We found that these
states are obtained for high input energies, whose value can also be due
uniquely to thermal photons, entering a BS of  extremely high
transmissivity, namely $\tau=1-\varepsilon$. Having unlimited thermal
resources at disposal, we can still
generate quantum correlated output states up to a value
$\mathcal{D}_{1|2}=1$, by sending a very excited input state in an
unbalanced BS of very high transmissivity.   If, always keeping the BS
transmissivity close to one, the fraction of squeezed photons is such to
render the input state $P$-nonclassical one, the corresponding points in
the plane will lie just below the separability threshold, but the value
of the discord can increase only up to the value $1-\log2\approx 0.3069$
(indicated by a red dashed line). By further increasing the amount of
squeezing, the resulting states will eventually occupy more entangled
and more discordant regions of the lower branch.  
%%%
\section{Nonclassicality arising from mixing a Gaussian state 
with a thermal state}
\label{ss:the}
We now consider the general case, allowing for thermal photons to enter
the second port of the BS.  In fact, in practical scenarios a certain
amount of thermal noise (\textit{e.g.}, in the form of black body
radiation or scattered light) unavoidably  participates in the
interference phenomenon and affects the statistics of the outgoing
fields.
\par
Though $P$-classicality, as already discussed, retains the expression
(\ref{NsNCl}) for the threshold value, the presence of another source of
photons affects the properties of the output state. The relevant
changes both to quantum discord and entanglement can be again evaluated
via the symplectic invariants, which now reads 
\begin{align}\label{symplinvth}
I_1=&\frac{1}{4}+n_2^2 (1-\tau )^2 +\tau\left[n_t
+n_s\left(1+2n_t\right)\left(1-\tau\right)+\tau \,n_t^2 \right]  \nonumber
\\ & \nonumber \\ &    +n_2 (1-\tau ) \left[1+2n_1\,\tau \right]  \, ,  \nonumber \\ & \nonumber \\
I_2=&\frac{1}{4}+n_t^2 (1-\tau )^2 +\tau\left[n_2
+n_s\left(1+2n_2\right)\left(1-\tau\right)+\tau \,n_2^2 \right]  \nonumber
\\ & \nonumber \\ &    +n_t (1-\tau ) \left[1+2\left(n_s+n_2+2n_sn_2
\right)\tau \right]  \, ,  \nonumber \\ & \\ I_3=&\left[n_2^2 + n_t^2 -
n_s(1+2 n_t)-2n_1n_2\right] (1-\tau) \tau \, ,\nonumber \\ & \nonumber
\\ I_4=&\frac{1}{16}\left(1+2 n_t \right)^2 \left(1+2 n_2 \right)^2 \, ,
\nonumber 
\end{align}
whence, recalling the expression for the total energy in mode $\a$ Eq. (\ref{toten}), we can see that 
$I_1$ and $I_2$ are each other related via the exchange of the number of thermal photons  $n_t$ 
and $n_2$.  
%%%%
\subsection{Generation of Gaussian discord}
%%%%%%%%%%%%%%%%%%%%%%%%%%%%%%%%%%%%%%%%%
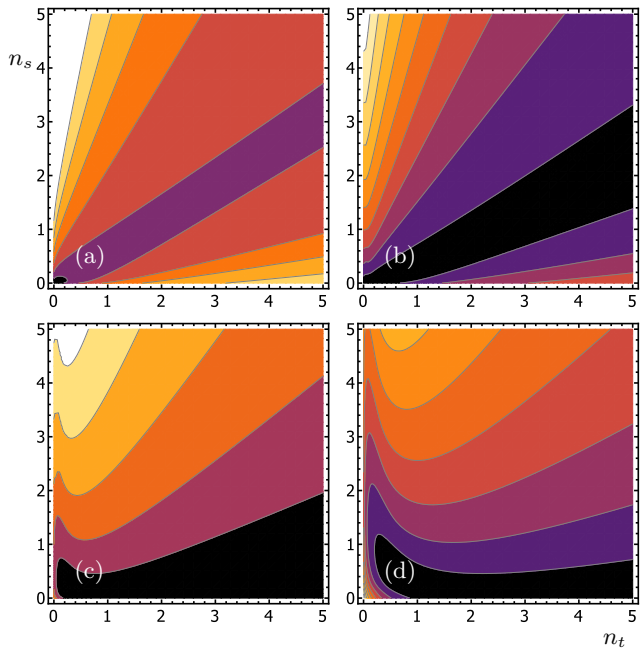
\begin{figure}
\subfigure{\label{N2_0}
\includegraphics[width=3.9cm]{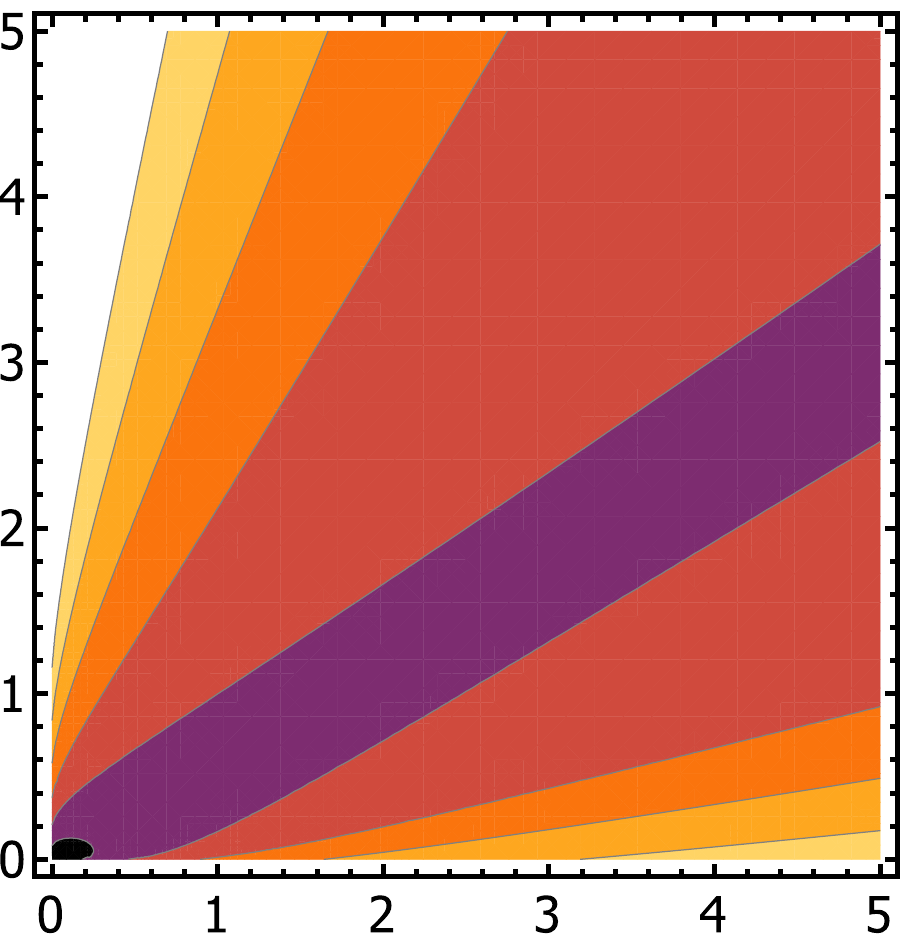}}
%\vskip -0.1cm
\subfigure{\label{N2_01}
\includegraphics[width=3.9cm]{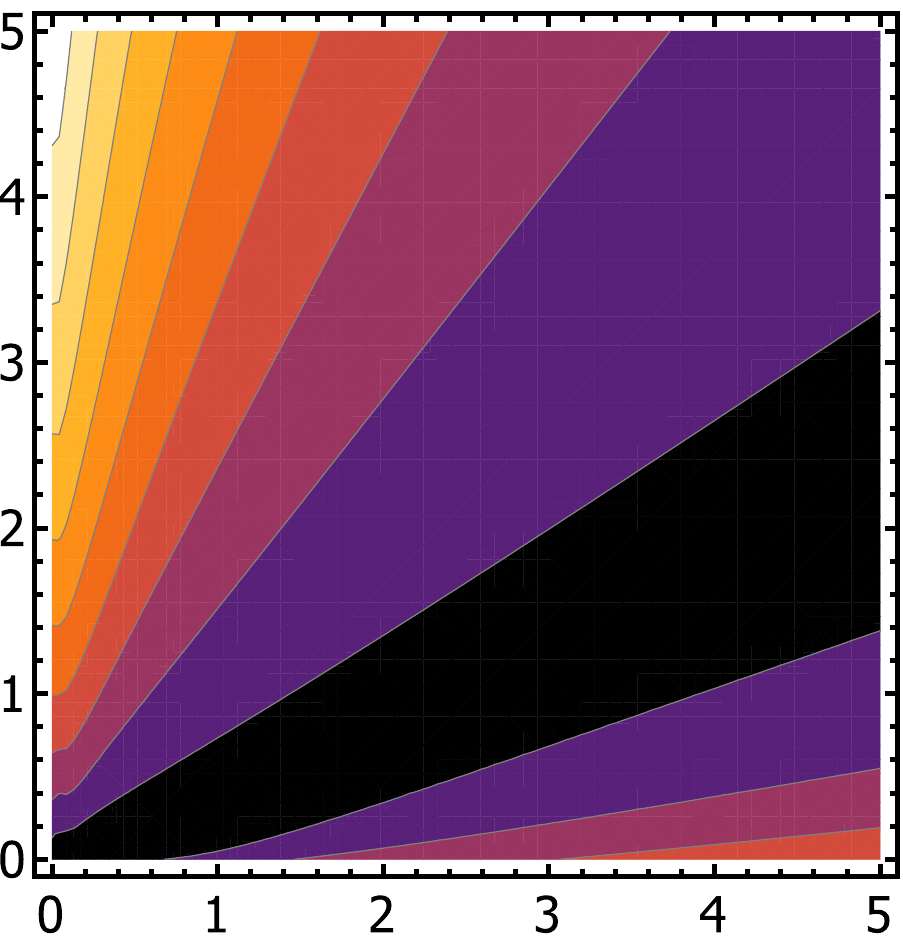}}
\begin{picture}(3,3)(117,52)
\put(-125,145){$n_s$}
\put(-100,70){\white (a)}
\put(17,70){\white (b)}
\end{picture}\\
\vskip -0.1cm
\subfigure{\label{N2_1}
\includegraphics[width=3.9cm]{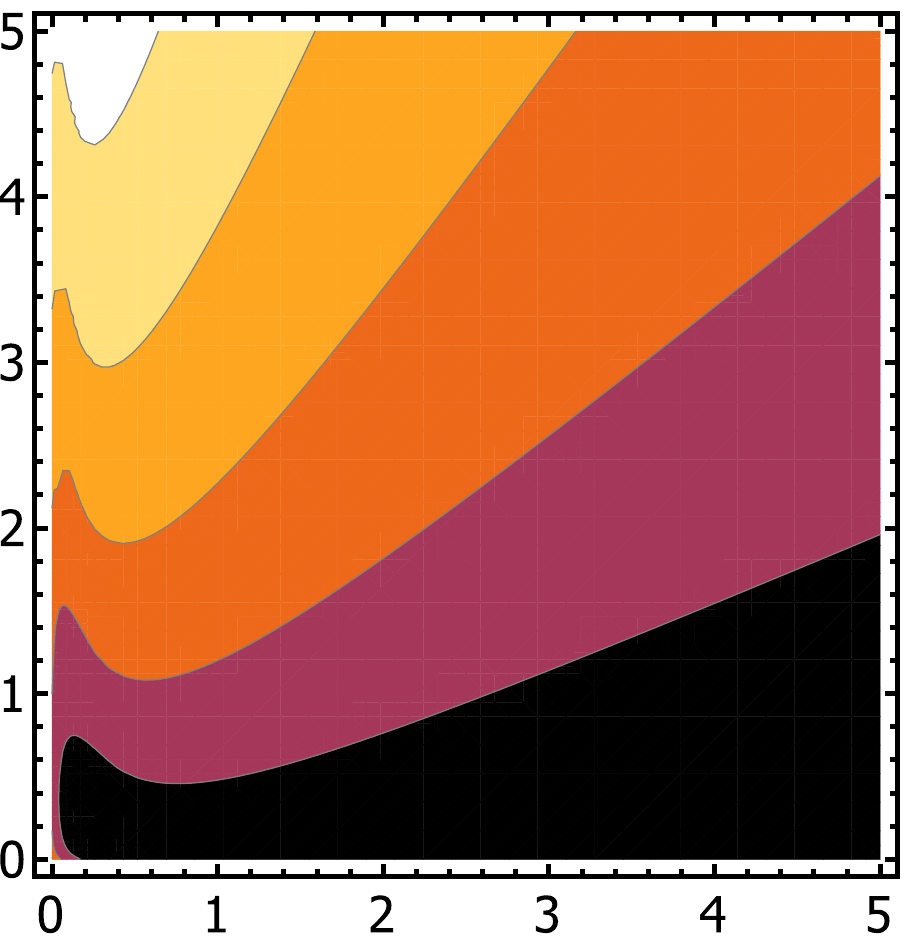}}
\subfigure{\label{N2_3}
\includegraphics[width=3.9cm]{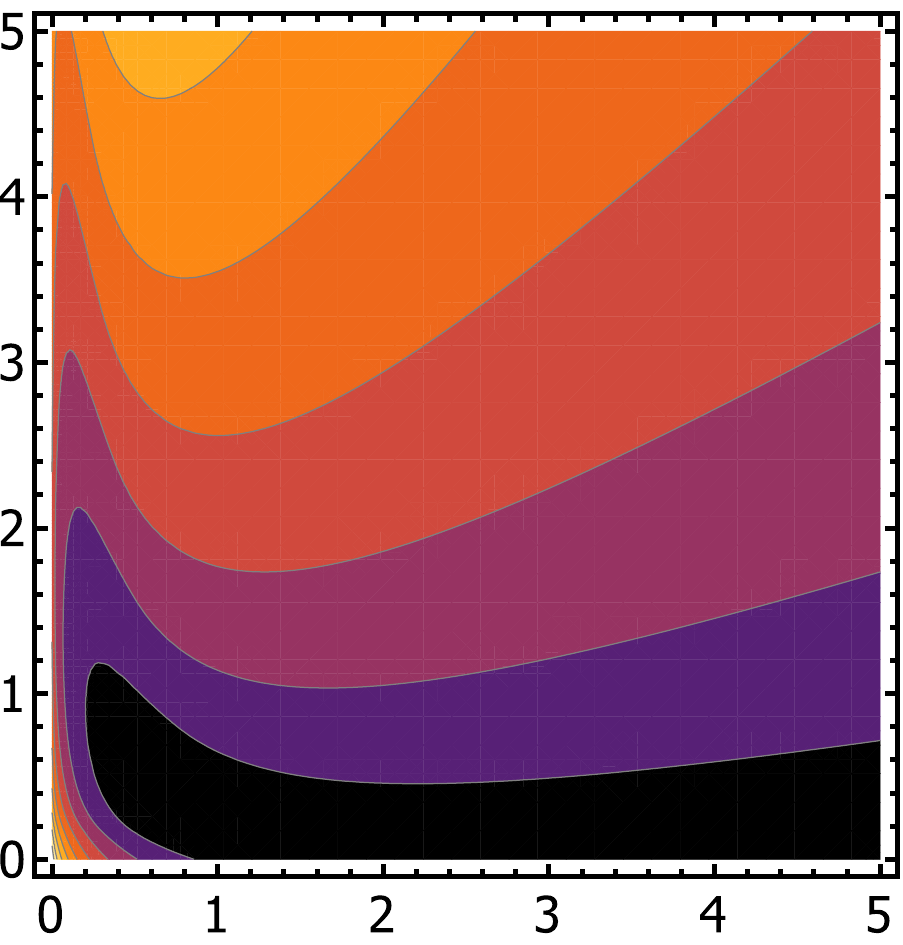}}
\begin{picture}(3,3)(117,52)
\put(-100,70){\white (c)}
\put(100,45){$n_t$}
\put(17,70){\white (d)}
\end{picture}
\caption{(Color online) 
Nonclassicality by mixing with a thermal state: 
contour plots of Gaussian discord as a function of 
squeezed and thermal photons $n_s$ and $n_t$ in mode $a_1$. From panel
(a) to (d) the number thermal photons in mode $a_2$ are given by
$n_2=0,0.1,1,3$ [panel (a) is in fact the contour plot of
Fig.~\ref{f:Discord3D}]. Colors go from black (0.) to white (0.7).}\label{f_ThermalD} \end{figure}
%%%%%%%%%%%%%%%%%%%%%%%%%%%%%%%%%%%%%%%%%
\par
As in the previous Section, we first focus on the case of a balanced
BS ($\tau=1/2$). In Fig.~\ref{f_ThermalD} we show the
contours of quantum discord as a function of $n_s$ and $n_t$ for
different thermal-photon number $n_2$ of mode $a_2$. We can see that,
despite the $P$-classicality of the output state remains invariant, the
$C$-classicality is much affected by the presence of an additional
source of thermal photons. In particular, for a low number of thermal
photons $n_2$ the region with minimal discord (darker areas in
Fig.~\ref{f_ThermalD}) localizes close to the $P$-classicality
threshold, whereas it tends to get closer to the zero squeezed-photon
axis for larger $n_2$. Also in this case, the inequivalence of the two
notions of non-classicality is apparent.
\par
\begin{figure}[h!] 
\centering 
\includegraphics[width=0.9\columnwidth]{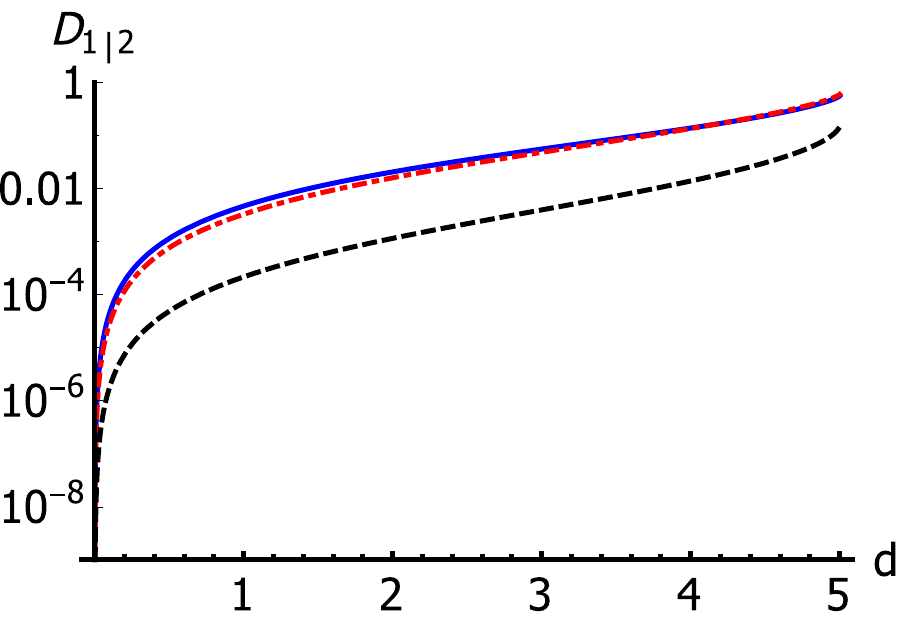}
\caption{(Color online) 
Nonclassicality by mixing with a thermal state: 
logarithmic plot of the $b_1$-discord $\mathcal{D}_{1|2}(R)$ as a
function of the imbalance $d$ for different values of the transmissivity
$\tau$ and fixed total energy $N=5$.  The solid blue curve is for a
balanced BS $\tau=0.5$, the dot-dashed red line is for $\tau=0.8$ while
the black dashed line is for $\tau=0.99$.}\label{f:DiscComparison}
\end{figure}

Since there is no threshold for the production of discord, it is
legitimate to enquire the generation of quantum correlations at a BS for
the ''cheapest" conceivable scenario, namely having at disposal only
\textit{thermal resources} in input.  Given a certain amount of total thermal
photons $N=n_1+n_2$, which is the most convenient redistribution of the
total energy between the two modes, in order to maximize the Gaussian
discord at the output?  The answer to this question is shown in In Fig.
\ref{f:DiscComparison}, where $b_1$-discord as a function of the photon
imbalance $d=n_1-n_2$ between the two input modes has been plotted. For
each transmissivity, the $b_1$-discord is a monotonically increasing
function of the imbalance $d$, so we can conclude that the optimal
configuration is the most asymmetric one, where all the thermal photons
are sent in on channel, leaving the other in the vacuum state.
Moreover, an even distribution of photons ($d=0$) between the input
modes always leads to zero output discord.  This fact is apparent by
looking at Eq. (\ref{sigmaeq}) where for equal input states, no matter
the transmissivity, the correlation terms $c_{\pm}$ of the CM
identically vanish: the phenomenon is referred to as transparency, since
the evolution trough the BS does not leave any imprint on the input
states. Being the optimal configuration the one with a thermal input in
one port of the BS and the vacuum in the other, we already know that,
for a given amount of energy $N$, there will be an optimal value of the
transmissivity maximizing the $b_1$-discord (as shown by the blue curve
of Fig. \ref{f:Disctau}); this fact is also manifest in the crossing of
the blue curve and the red dot-dashed one in Fig.
\ref{f:DiscComparison}, corresponding to $\tau=0.5$ and $\tau=0.8$
respectively, when approaching the maximum imbalance. Finally, the
dashed curve corresponds to an extremely unbalanced BS, namely
$\tau=0.99$, and, provided that high-enough thermal energy is available,
the corresponding value of the $b_1$-discord close to the imbalance
would be the greatest, eventually achieving the limiting value of
$\mathcal{D}_{1|2}=1$, as discussed above for the circled points in Fig.
\ref{f:RndDiscVac}.    
%%%
%%%%
\subsection{Generation of Gaussian entanglement}
As before, the equation $\tilde{\lambda}(n_s,n_t,n_2,\tau)=1/2$, if
solved with respect to the squeezed number of photon $n_s$, gives a
threshold on the generation of entanglement when $n_2$ thermal photons
enter the BS. 
%%%
\begin{figure}[h!] 
\centering 
\includegraphics[width=0.9\columnwidth]{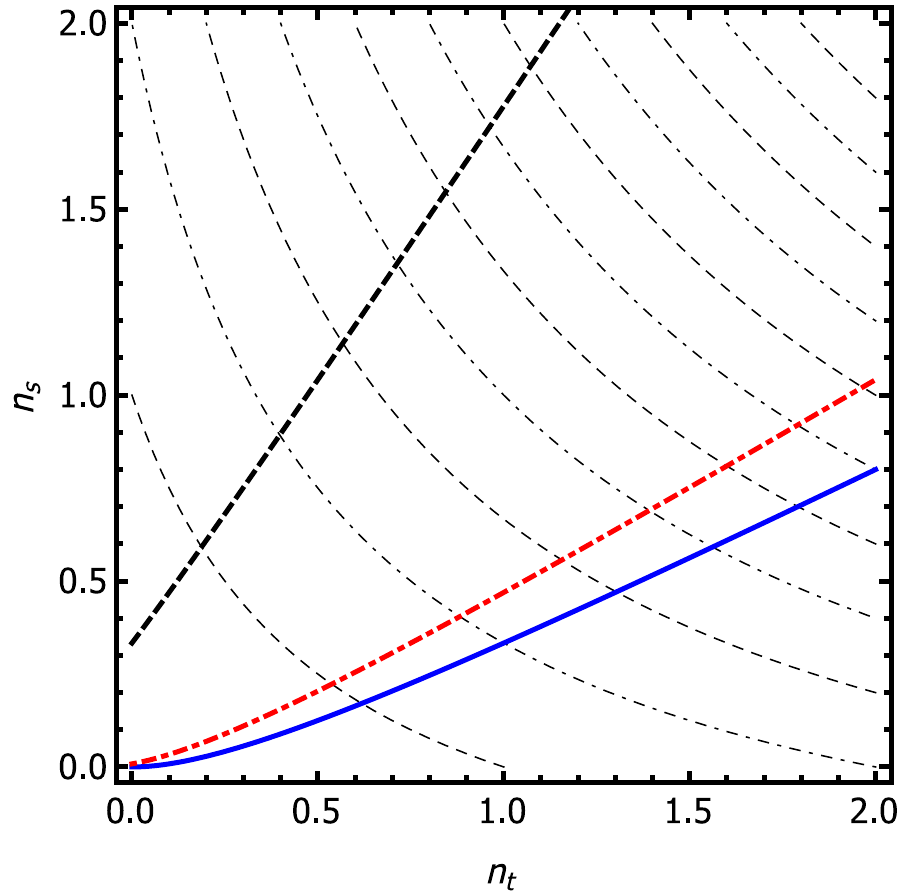}
\caption{(Color online) 
Nonclassicality by mixing with a thermal state: 
plot of the separability thresholds $n_s^{\mathsf{sep}}$ for
fixed $\tau=\frac12$ and different values of $n_2$.  The solid blue line
corresponds to $n_2=0$, and indeed coincides with the non-classicality
threshold $n_s^{\mathsf{nc}}$.  The red dot-dashed lines represents
$n_s^{\mathsf{sep}}$ for $n_2=0.1$, and finally the black dashed
one is for $n_2=1$. The black lines are curves of fixed energy in the
mode $\a$ $n_1=n_s +n_t+2n_sn_t$ : dashed for odd values $n_1=2k+1\, ,\,
k=0,1,\ldots$ and dot-dashed for even values $n_1=2k ,\, k=1,2,\ldots$
of the total energy.\label{f:SepTh}} \end{figure}
The explicit expression of $n_s^{\mathsf{sep}}$ is given by   
\begin{equation}\label{NsSepth}
n_s^{\mathsf{sep}}=\frac{\mu_1\, \mu_2}{\tau(1-\tau )} \Theta_{t,2}\,\Theta_{2,t}  \; ,
\end{equation}
where  $\Theta_{k,l}=n_k n_l +n_k-( n_k-n_l)\tau$ and $\mu_{1,2}$ are the purities
of the two input states.
\par
\begin{figure}[h!] 
\centering 
\includegraphics[width=0.9\columnwidth]{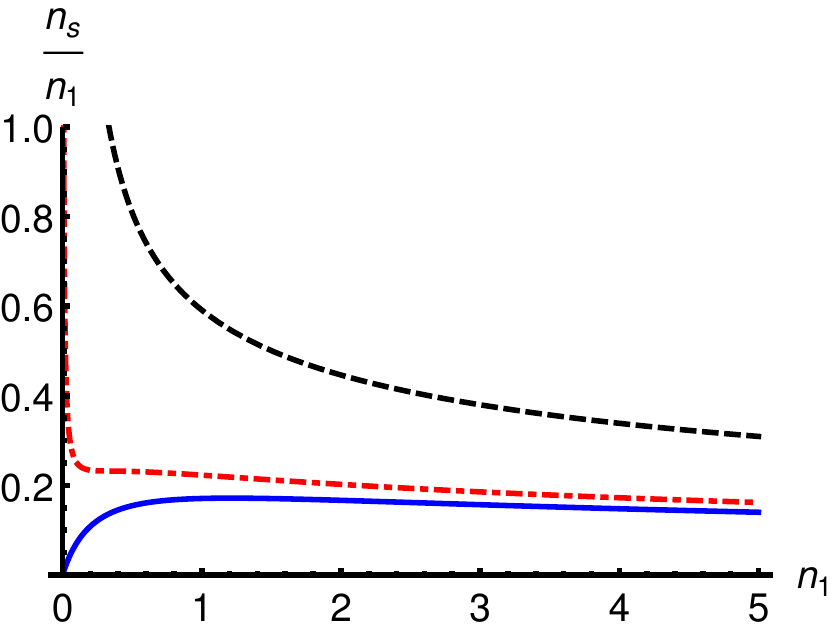}
\caption{(Color online) 
Nonclassicality by mixing with a thermal state: 
plot of the squeezed fraction of photons $n_s/n_1$ at the
separability threshold, as a function of the total number of photons
$n_1=n_s+n_t+2n_sn_t$ entering the first port of the beam splitter, for
different values of $n_2$.The solid blue line corresponds to $n_2=0$,
The red dot-dashed line corresponds to $n_2=0.1$, while the black dashed
one to $n_2=1$.
\label{f:FracNsNthTh}}
\end{figure}

Contrary to the vacuum case, it is apparent that \STErev{in the presence
of a thermal state} the separability threshold $n_s^{\mathsf{sep}}$ and
the non-classicality threshold $n_s^{\mathsf{nc}}$ are no longer
coincident. In Fig.~\ref{f:SepTh} several separability thresholds are
shown, for different values of $n_2$. As soon as $n_2$ differs from
zero, the number of squeezed photons $n_s$ required to have entanglement
increases --- as shown both in Fig.~\ref{f:SepTh} and
Fig.~\ref{f:FracNsNthTh}.  
The previous symmetry between the notions of non-classicality in the
phase space and non-separability no longer holds: there exists
$P$-singular input states of the electromagnetic field, and hence
$P$-singular output states, which nevertheless are not entangled.  We
can conclude that a hierarchy of non-classicality has settled down:
non-separability at the output imposes a stricter notion of quantumness
than the one put forward by $P$-singular distributions. Injecting into
the BS \textit{a non-classical state is no longer a sufficient
condition to get entanglement between the output modes}.  
\par 
\STErev{In order to better investigate this point,} we express the
separability threshold relative to $n_2$ thermal photons as a function
of the $P$-nonclassicality threshold $n_s^{\mathsf{P}}$. We focus on
the optimal case of a balanced BS, obtaining:
\begin{equation}
n_s^{\mathsf{sep}}=\frac{\left[\,n_2+h(n_s^{
\mathsf{P}})(1+2n_2)\,\right]^2}{(1+2n_2)[1+2\,h(n_s^{\mathsf{P}})\,]} \, ,
\end{equation}
where
$h(n_s^{\mathsf{P}})=n_s^{\mathsf{P}}+
\sqrt{n_s^{\mathsf{P}}(1+n_s^{\mathsf{P}})}$
is a monotonically increasing function of the non-classicality
threshold.  Even if the two thresholds $n_s^{\mathsf{sep}}$ and
$n_s^{\mathsf{P}}$
now differ, their knowledge enables one, given a known amount of thermal
noise in $\b$, to estimate the effective $P$-nonclassicality required in
$\a$, i.e. how much squeezing pump into the BS, in order to get
entanglement.  When we have a squeezed vacuum state entering the BS in
the mode $\a$, namely $n_t=0$, and $n_2$ thermal photons in $\hat{b}$,
the separability threshold in Eq.~(\ref{NsSepth}) reduces to
$n_2^2/(1+2n_2)$, independently on $\tau$. It is the value of the curves
$n_s^{\mathsf{sep}}$ at $n_t=0$, as it can be seen from
Fig.~\ref{f:SepTh}, and moreover it is the same expression as
$n_s^{\mathsf{nc}}$ with  $n_t$ replaced by $n_2$. Thus, having a
squeezed vacuum state in mode $\a$ and a thermal state in mode $\b$
(characterized by $n_s$ squeezed and $n_2$ thermal photons,
respectively) is equivalent to have a single mode Gaussian state
$\varrho(n_s,n_2)$ in $\a$ and the vacuum in $\b$.
\par
\begin{figure}[h!] 
\centering 
\includegraphics[width=0.9\columnwidth]{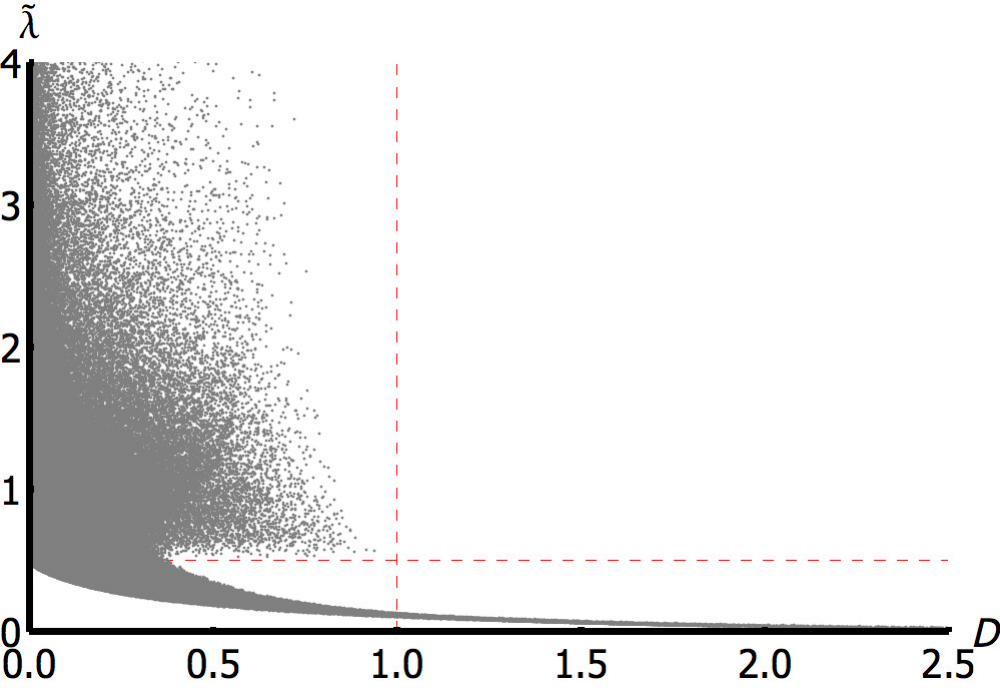}
\caption{(Color online) 
Nonclassicality by mixing with a thermal state: 
symplectic eigenvalue of the partial transpose
$\tilde{\lambda}_-$ versus discord for randomly generated
input states $\varrho(n_s,n_t)$, random values of the transmissivity
$\tau$ and randomly generated thermal states $\nu(n_2)$ at the second
port of the beam splitter. \label{f:RndDiscTh}}
\end{figure}

Finally, in Fig.~\ref{f:RndDiscTh} we propose the same random plot
as in Fig.~\ref{f:RndDiscVac}, with the difference that random number of
thermal photons is added in the second mode. We can see that the lower
branch is substantially unchanged by the presence of thermal noise; even
if the entanglement sets later, i.e. for higher amount of squeezing, the
relationship with Gaussian discord remains the same. On the other hand,
in the remaining accessible region of the plane, compared with
Fig.~\ref{f:RndDiscVac}, the points are scattered and the sharp pattern
is now washed out. On average the distribution drops towards lower
values of the discord. 
\subsection{Effective non-classicality and non-classical depth}
As said above, in the case in which thermal photons are injected in the
second port of the BS, the $P$-non-classicality is no longer a necessary
and sufficient condition to obtain output entanglement. However,
remarkably, a quantitative relation between these two notions can still
be worked out. In particular, we will now see that the non-classical
depth at the input determines the potential of generating entanglement
at the output.  
\par
Let us first consider the implicit equation defining the separability
threshold $\tilde{\lambda}_-(n_s,n_t,n_2,\tau)=1/2$, and let us call
$\mathcal{E}_{\varrho}(\tau)$ its solution with respect to the number of
thermal photons in the second port \STErev{as a function of $\tau$}.  It
expresses (as a function of the input parameters $n_s, n_t$) the number
of thermal photons that can enter a BS of transmissivity $\tau$ in the
$\b$ mode and yield an output entangled state. The explicitly form of
$\mathcal{E}_{\varrho}(\tau)$, although analytical, is quite cumbersome
and hence has not been reported. If we now perform a maximization over
the transmissivity $\tau$ we obtain the following quantity:
\begin{equation}\label{effective}
\mathcal{E}_{\varrho}=\max_{\tau}\,\mathcal{E}_{\varrho}(\tau) \, .
\end{equation}
We shall refer to this quantity as to the\textit{ effective
non-classicality} of the state $\varrho$ and it embodies the maximum
allowed number of thermal photons that can be mixed with $\varrho$ at a
BS and still get an entangled output state. 
\par
While the non-classical depth is a property of a single-mode state of
the field, the effective non-classicality is a property of the two-mode
configuration that we are considering.  In other words, the effective
non-classicality $\mathcal{E}_{\varrho}$ must be intended as an attempt
to characterize \textit{operationally} the non-classical feature of a
state. Thus, the relation between $\mathcal{E}_{\varrho}$ and $\ncd_m$,
if any, is a priori unclear. Let us stress that the operational
interpretation commonly associated to the $\ncd_m$ of a single-mode
state $\varrho$ is that it gives the number of thermal photons that have
to be statistically mixed with the state $\varrho$ in order to obtain a
classical state. In this sense, this operational interpretation of
$\ncd_m$ exclusively refers to single-mode states. 
\par
In order to clarify the relation between $\mathcal{E}_{\varrho}$ and
$\ncd_m$ we notice first that, by means of a numerical maximization, it
is possible to show that $\mathcal{E}_{\varrho}$ is always obtained for
$\tau=1/2$.  Thus the balanced BS represents the overall optimal
configuration, and in this case the effective non-classicality reads  
\begin{equation}
\mathcal{E}_{\varrho}=\frac{n_s - n_t+ \sqrt{n_s (1 + n_s)} }{1 + 2 n_t}
\, .  
\end{equation} 
If now we look at the expression of the
non-classical depth Eq. (\ref{NCdepthmax}) and insert it in
$\mathcal{E}_{\varrho}$, after some manipulation we find the following
relation
\begin{equation}\label{enc_ncd}
\mathcal{E}_{\varrho}=\frac{\ncd_m}{1-2\ncd_m} \, .
\end{equation}
Thus, the two quantities $\mathcal{E}_{\varrho}$ and $\ncd_m$ which, as
said, are defined in reference to different systems, are in fact related
via a simple expression. In other words, this endows the non-classical
depth with a new operational interpretation: \textit{the non-classical
depth of a state determines, via Eq.~(\ref{enc_ncd}), the maximum number
of thermal photons that can be mixed with it at a beam
splitter without destroying the output entanglement}.
%%%
\section{Conclusions}
\label{s:out}
The quantum-to-classical transition for a single-mode bosonic system may
be fully characterized by the properties of its Glauber-Sudarshan
$P$-function in the phase space. On the other hand, for two-mode states,
quantumness may be recognized either by the presence of quantum
correlations ($C$-nonclassicality) {\em or} in terms of its phase space
distribution ($P$-nonclassicality).  In this paper we have addressed the
generation of both types of nonclassicality by the linear mixing of a
single-mode Gaussian state with a thermal state at a beam splitter, and
have explored in details the relationships between the nonclassical
features of the single-mode input and the $P$- and $C$-nonclassicality
of the two-mode outputs. 
\par
We have shown that, for mixing with vacuum, a balanced BS is capable of
generating $C$-nonclassicality for any input state, contrary to the case
of $P$-nonclassicality. In addition, the $C$-nonclassicality increases
as the input $P$-nonclassicalitity decreases. These findings clearly
confirm in a dynamical setting the inequivalence between these two
notions of nonclassicality that was highlighted in Ref.~\cite{PvsI} in a
geometrical context. We confirm this inequivalence also for mixing with a
thermal state, even if more complex behaviors emerge.  
\par 
In addition, we have shown that input $P$-classicality and output
separability single out two thresholds which coincide only for the case
of linear mixing with the vacuum, whereas they are connected in a non
trivial way for linear mixing with a thermal state. In fact,
$P$-classicality at the input, as quantified by the non-classical depth,
does determine quantitatively the potential of generating output
entanglement.  This allows us to provide a new operational
interpretation for the non-classical depth: it gives the maximum number
of thermal reference photons that can be mixed at a beam splitter
without destroying the
output entanglement. 
\par
By reinforcing quantitatively the inequivalence between $P$- and
$C$-classicality, our results paves the way for analyzing the dynamical
relationship between different types of non-classicality in more
general contexts.
\acknowledgments
This work has been supported by MIUR through the FIRB project
``LiCHIS'' (grant RBFR10YQ3H), by EU through the Collaborative 
Projects TherMiQ (Grant Agreement 618074) and QuProCS (Grant 
Agreement 641277) and by UniMI through the H2020 Transition 
Grant 14-6-3008000-625.
%%%%

%%%
\end{document}